\documentclass[12pt]{article}
\usepackage[utf8]{inputenc}
\usepackage{amsmath,graphicx}

\usepackage[square,numbers,sort&compress]{natbib}
\usepackage{multicol,multirow}
\usepackage{fullpage}
\usepackage{lineno}
\usepackage{xcolor}
\usepackage{pdflscape}
\usepackage{hyperref}
\usepackage{authblk}

\newcommand{\highlight}{}

\title{Quantifying age-specific household contacts in Aotearoa New Zealand for infectious disease modelling}

\author[1]{Caleb Sullivan}
\author[1,2]{Pubudu Senanayake}
\author[1,*]{Michael J. Plank}

\affil[1]{School of Mathematics and Statistics, University of Canterbury, Christchurch, New Zealand}
\affil[2]{Stats NZ, Christchurch, New Zealand}

\affil[*]{Corresponding author: Michael J. Plank, michael.plank@canterbury.ac.nz}

\date{}

\begin{document}


\maketitle

\begin{abstract}
Accounting for population age structure and age-specific contact patterns is crucial for accurate modelling of human infectious disease dynamics and impact. A common approach is to use contact matrices, which estimate the number of contacts between individuals of different ages. These contact matrices are frequently based on data collected from populations with very different demographic and socioeconomic characteristics from the population of interest. Here we use a comprehensive household composition dataset based on Aotearoa New Zealand census and administrative data to construct a household contact matrix and a synthetic population that can be used for modelling. We investigate the behaviour of a compartment-based and an agent-based epidemic model parameterised using this data, compared to a commonly used contact matrix {\highlight that was constructed by projecting international data onto New Zealand's population}. We find that using the New Zealand household data, {\highlight either in a compartment-based model or in an agent-based model}, leads to lower attack rates in older age groups compared to using the {\highlight projected} contact matrix. This difference becomes larger when household transmission is more dominant relative to non-household transmission. We provide electronic versions of the synthetic population and household contact matrix for other researchers to use in infectious disease models. 
\end{abstract}


\newpage

\section{Introduction}

Age is a crucial variable affecting human infectious disease dynamics and impact. Many pathogens have strong age gradients in clinical severity, meaning that health impact and demand for healthcare are highly dependent on the age distribution of infections, e.g. SARS-CoV-2 \cite{verity2020estimates}, influenza \cite{mcdonald2023inference}. The immune response to infection or vaccination can be age-dependent, which has implications for disease transmission dynamics \cite{castle2000clinical,muller2021age}. For endemic diseases that confer lasting immunity and for childhood vaccination diseases, susceptible individuals are concentrated in the youngest age groups who have not yet been infected or vaccinated, e.g. measles, pertussis \cite{anderson1985vaccination}. Variations in contact rates with age mean that transmission rates are unequal, and this tends to change the overall attack rate relative to a well-mixed population \cite{delvalle2013mathematical}. 

Households play a key role in infectious disease dynamics, as they comprise relatively static groups of individuals who typically spend relatively large amounts of time in close contact \cite{riley2007large}. For many infectious pathogens, a significant amount of transmission occurs within households \cite{cauchemez2004bayesian,kwok2011modelling,paul2021characteristics}. Average household size and household structure can have a major influence on infectious disease dynamics \cite{house2009household,hilton2019incorporating}.

Infectious disease models therefore need to account for population age structure and age-specific contact patterns within and outside households. A standard 
approach to including age structure in compartment-based epidemic models is to divide the population into age groups \cite{keeling2011modeling}. This requires some parameter estimates for the contact rates {\highlight (i.e. the average number of contacts between pairs of individuals per unit time that could potentially result in disease transmission)} between individuals in different age groups. This is typically expressed as a contact matrix or next generation matrix \cite{schenzle1984age,diekmann2010construction}.

The gold standard method for estimating contact matrices is diary-based contact surveys, in which a sample of individuals record how many contacts of a given age they had over a defined time period in different household and non-household settings. However, conducting diary-based surveys on a representative sample is costly and challenging, and such studies have been conducted infrequently. One of the most frequently used studies is the POLYMOD study \cite{mossong2008social}, in which 7,290 participants across 8 European Union countries recorded the age, gender, and other aspects of their contacts in a 24-hour period in 2005-2006. 

Prem et al. \cite{prem2017projecting} used a Bayesian hierarchical model to project the contact patterns in the POLYMOD study and create synthetic contact matrices for 152 countries, with contacts divided into home, school, work and other settings. To account for differences between countries in household structure and contact patterns, nine demographic and socioeconomic indicators were used to weight the results from the POLYMOD countries. Prem et al. updated their results and produced synthetic contact matrices for 177 geographical regions in 2021 \cite{prem2021projecting}. 

Because direct empirical data on contact patterns are rarely available, synthetic contact matrices are frequently needed \cite{conmat}. The synthetic contact matrices estimated by Prem et al. \cite{prem2017projecting,prem2021projecting} have been used extensively in the infectious disease modelling literature, with almost 100 citations per year on PubMed between 2020 and 2023. Many models of Covid-19 for policy advice \cite{moore2021modelling,bubar2021model,ram_schaposnik_2021,nguyen2021covid,kimathi2021age,conway2023covid} have used country-specific contact matrices based on these estimates. During the Covid-19 pandemic, some studies used real-time survey data to estimate the effect social distancing measures and behaviours were having on age-specific contact rates \cite{gimma2022changes,golding2023modelling}. However, this data is not routinely available and there is a need for data and methods that can be used to estimate baseline (i.e. non-pandemic) contact patterns with data that is representative of the population of interest.

In Aotearoa New Zealand, age-structured epidemic models were used during the Covid-19 pandemic to inform the government's strategy and public health response \cite{vattiato2022assessment,lustig2023modelling,datta2024impact}. These models used contact matrices estimated by Prem et al. \cite{prem2017projecting} for New Zealand, with some adjustments made to account for New Zealand's 2021 population age structure \cite{steyn2022covid}. However, these matrices were derived from social survey data collected in European countries in 2005-06, {\highlight and the only New Zealand-specific data used was population age structure. Therefore, they may not accurately reflect contemporary age-specific contact patterns in New Zealand. In particular, household composition patterns in New Zealand are likely to be different to those in POLYMOD countries. Furthermore, New Zealand's age structure and household composition patterns will change over time and so methods are needed to enable contact matrices to be updated using more recent, locally sourced data. }

While there is no New Zealand-specific contact pattern data across all settings, household composition data may be used to obtain a more accurate description of contact patterns within households. In this study, we use official New Zealand census and administrative data on household composition in 2018 to derive a New Zealand-specific household contact matrix. The data represents an estimated 89\% of the total population, so while not complete, does include the large majority of individuals. We compare the contact matrix constructed from this data with the home contact matrix estimated by Prem et al. \cite{prem2017projecting} for New Zealand. We also compare the results of simple age-structured epidemic models using the different matrices, and an agent-based model that accounts for household structure explicitly. 

{\highlight There are three key benefits to our approach relative to that of \cite{prem2017projecting}. Firstly it provides a household contact matrix that is based on census rather than survey data. Secondly, it uses New Zealand-specific data on household composition rather than taking a different country’s household age matrix and projecting it onto New Zealand’s age structure. Thirdly, it uses more recent data and provides a reproducible method for updating the results as updated data becomes available from future censuses. }

We provide a public repository containing the raw household composition data, the code used to analyse the data, and the household contact matrix and synthetic population that this produces. These outputs may be useful for future efforts to model age-structured infectious disease dynamics in New Zealand. The algorithms and code may also be useful to researchers in other countries where comparable household composition data is available.

\section{Methods}
\subsection{Data} \label{sec:data}

Data on household composition were extracted from the 2018 New Zealand Census dataset.
The census in New Zealand collects information about dwellings and individuals, including information about their usual residence \cite{stats_nz_2018-census-population-and-dwelling-counts}. Because not everyone returns a census form, Stats NZ use administrative data (such as information from tax and birth records) to supplement the census information. 

In the 2018 census, 11\% of the population was enumerated using administrative data \cite{stats_nz_adding-admin-records}, but 357,294 individuals (7.6\% of the population) could not be confidently placed into dwellings. This resulted in responding dwellings with incomplete households (estimated to be 6.6\% of all households) and non-responding dwellings with no household information (3.6\% of all households). Overall, 3.0\% of households were either fully or partially sourced from administrative data \cite{stats_nz_2018_census_families_and_households}.

Using the individual attribute information to ascertain the ages of people in a dwelling, the various 10-year age-group combinations that occurred in the data were constructed and aggregated to provide counts of households comprising the given age-group combinations (see Table \ref{tab:Stats NZ data}a for an example). Individuals without specific dwelling information were excluded (approximately 8\% of individuals). To avoid disclosure, random rounding to base 3 and suppression of household compositions with counts of less than 6 were applied.  The construction of the data was conducted by Stats NZ’s Customised Data Services and provided by a Stats NZ customised report. 

To estimate the number of missing individuals in each age group, we compared to the 2018 census usually resident population \cite{stats_nz_2019}, which we will refer to as the census population (see Supplementary Table S3).

\begin{table}
    \centering
    \begin{tabular}{ccccccccc}
    \hline
    \multicolumn{8}{c}{{\bf Number of people in household by age group}} & {\bf Number of households}  \\
         0-9 & 10-19&20-29&30-39&40-49&50-59&60-69&70+& \\ \hline
        2 & 0& 0 & 1& 1& 0 & 0 & 0  &  13212 \\
        \hline       
    \end{tabular}
    \\
    \vspace{2mm}
    (a)

    \vspace{1cm}

     \begin{tabular}{lll}
    \hline
    Individual ID number & Age group & Household ID number \\
    \hline
      1 & 1 & 1 \\ 
      2 & 1 & 1 \\
      3 & 4 & 1 \\
      4 & 5 & 1 \\
       \hline
    \end{tabular}   
    \\
   \vspace{2mm}
     (b)
    \caption{(a) Example of a line from the raw Stats NZ household composition data showing a single household type $k$ showing that there were 13212 households in New Zealand that consisted of two 0-9-year-olds, one 30-39-year-old, and one 40-49-year-old. See Supplementary Table S1 for complete dataset. (b) Part of the synthetic population corresponding to a single household of the type shown in (a). This consists of four individuals (two in age group $1$, one in age group $4$ and one in age group $5$), all with the same household ID number. In this example, these four rows would be repeated $13212$ times, with each block of four rows having a new household ID number, to give individual ID numbers $1$ to $52848$ and household ID numbers $1$ to $13212$.  } 
    \label{tab:Stats NZ data}
\end{table}

\subsection{Imputing missing data} \label{sec:imputing}
A household type $k$ is defined by the number of people $H_{ki}$ in age group $i$ in a single household of type $k$ and we use $f_k$ to denote the total number of households of type $k$ according to the Stats NZ household composition data.

The degree of under-representation in the household composition data relative to the census population differed by age group. This meant that the raw data had a different age distribution to the real population. This will tend to bias the outputs of infectious disease models.  

\begin{figure}
    \centering
    \includegraphics[width=0.75\linewidth, trim={6cm 0.8cm 6cm 1cm},clip]{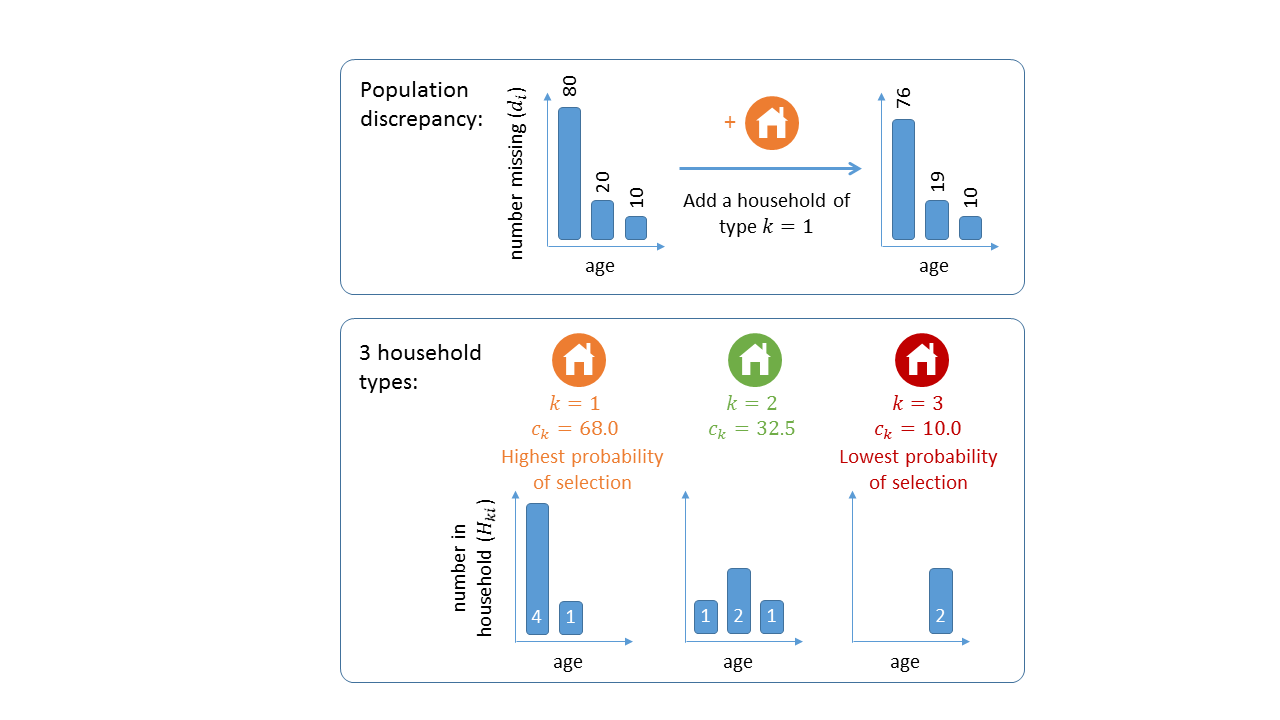}
    \caption{Schematic illustration of the imputation procedure in a simplified example with three age groups ($i=1,2,3$) and three household types ($k=1,2,3$). The initial discrepancy between the target population and the household population is $d_i=[80,20,10]$ in the youngest, middle and oldest age group respectively (top left). The three households types contain $H_{1i}=[4,1,0]$, $H_{2i}=[1,2,1]$ and $H_{3i}=[0,0,2]$ people in the three age groups, and so receive weights of $c_1=68.0$, $c_2=32.5$ and $c_3=10.0$ (bottom). Thus household type 1 is most likely to be selected to add to the population because its age structure is closest to that of the discrepancy vector $d_i$. This reduces the discrepancy (top right) and the process is then repeated.   }
    \label{fig:diagram}
\end{figure}

To address this, we applied an imputation method to adjust the frequency $f_k$ of each household type. This method used the following algorithm (see Figure \ref{fig:diagram}):
\begin{enumerate}
    \item Define $N_i=\sum_k H_{ki}f_k$ to be the population size in age group $i$ according to the household composition data and $N^\mathrm{targ}_i$ to be the census population.
    \item Set $d_i=N^\mathrm{targ}_i-N_i$ to be the discrepancy between these.
    \item For each household type $k$, calculate $c_k = \mathrm{max}\left(0, \sum_i H_{ki} d_i/\sum_i H_{ki}\right)$. Household types with high $c_k$ have an age distribution similar to the discrepancy vector ${\bf d}$, whereas those with low $c_k$ have an age distribution dissimilar to ${\bf d}$. {\highlight Note that the $\mathrm{max}(.)$ in the definition of $c_k$ prevents negative values being assigned for occasional instances where $d_i<0$ for some $i$.}
    \item Randomly select a household type $k$ with probability $c_k f_k^\mathrm{raw}/\sum_l c_l f_l^\mathrm{raw}$, where $f_k^\mathrm{raw}$ is the number of households of type $k$ in the raw data. Increase $f_k$ by 1.
    \item Repeat until the norm of discrepancy vector ${\bf d}$ is less than 1\% of the norm of the target vector ${\bf N}^\mathrm{targ}$.
    \end{enumerate}

We designed this algorithm to make the minimal necessary adjustments to the household frequencies $f_k$ to obtain an age-specific population that is sufficiently close to the census population. However, there is no unique way to impute the missing data and other imputation methods are possible. In particular, the imputation algorithm we used is restricted to adding households of a type already in the dataset with a minimum count of 6. In reality, there will be missing households of different types, suppression of household types with a small count, and misclassification of households as the incorrect type, e.g. due to individuals not being assigned to the correct household. These are limitations of the raw data.

\subsection{Deriving a contact matrix from household composition data} \label{sec:contact_matrix}

The total number of people in age group $i$ who were represented in the household composition data after imputation was $N_i = \sum_k H_{ki} f_k$.
We constructed a contact matrix $C_{ij}$ representing the average number of household contacts that an individual in age group $i$ has with individuals in age group $j$. This was calculated from the household composition data after imputation as:
\begin{equation} \label{eq:contact_matrix}
    C_{ij} = \frac{1}{N_i} \sum_{k} H_{ki} \left(H_{kj}-\delta_{ij}\right) f_{k},
\end{equation}
where $\delta_{ij}$ is the Kronecker delta function. Subtracting $\delta_{ij}$ meant that self-contacts were excluded from the count. The value of the summation is the total number of contacts between individuals in age groups $i$ and $j$. Dividing this by the population size $N_i$ in age group $i$ gave the average number of household contacts per person. 

{\highlight Note that the matrix $C_{ij}$ defined by Eq. \eqref{eq:contact_matrix} corresponds to what Prem et al. \cite{prem2017projecting} referred to as the ``household age matrix'' (HAM). The home contact matrix constructed by Prem et al. is a different matrix in that: (i) it includes contacts that occurred in the home with visitors; and (ii) it uses POLYMOD data to map from the number of cohabitants in the HAM to the number of the people with whom contact was reported in the 24 hour period of the diary study. Here, we make the simplifying assumption that individuals have contact with all of their household members during a typical infectious period (noting that contact may or may not lead to transmission depending on the transmission rate parameter in the model). Our definition of $C_{ij}$ also implicitly categorises contacts with visitors as non-household contacts. This is reasonable as, although the contact may have occurred within a home, it leads to the possibility of between-household transmission in the event that the visitor(s) returns to their own home during their infectious period.  }

\subsection{Balancing contact matrices} \label{sec:balancing}
Since the total number of contacts between individuals in groups $i$ and $j$ is the same as the number of contacts between individuals in groups $j$ and $i$, any contact matrix $C_{ij}$ should satisfy the balance equation \cite{arregui2018projecting}:
\begin{equation} \label{eq:detbal}
N_i C_{ij} = N_j C_{ji}.
\end{equation}
The contact matrix constructed from the household composition data satisfied this condition {\highlight with respect to the imputed population size vector $N_i$} by construction. However, the contact matrices estimated by Prem et al.  \cite{prem2017projecting} for New Zealand did not satisfy this condition for the New Zealand population (or in fact for any age distribution $N_i$). This tends to skew the results of epidemic models using these contact matrices \cite{hamilton2024examining}, particularly if there is a high degree of mismatch in Eq. (\ref{eq:detbal}) for the population being modelled. 

We therefore forced the contact matrices of \cite{prem2017projecting} to satisfy Eq. (\ref{eq:detbal}) by defining a modified matrix $C$ from the original matrix $\tilde{C}$ \cite{arregui2018projecting,steyn2022covid}: 
\begin{equation}
C_{ij} = \frac{1}{2}\left( \tilde{C}_{ij} + \frac{N_j}{N_i} \tilde{C}_{ji}\right).
\end{equation}

\subsection{Compartment-based epidemic model} \label{sec:ODE_model}
We modelled epidemic dynamics using a simple compartment-based model for the number of susceptible ($S_i$), exposed ($E_i$), infectious ($I_i$) and recovered ($R_i$) individuals in each group $i$ assuming a closed population and permanent immunity to reinfection. This was described by the following standard system of ordinary differential equations \cite{diekmann2000mathematical}:
\begin{eqnarray}
\frac{dS_i}{dt} &=& -\lambda_i S_i,  \\
\frac{dE_i}{dt} &=& \lambda_i S_i - \gamma E_i, \\
\frac{dI_i}{dt} &=& \gamma E_i - \mu I_i,
\end{eqnarray}
where $\lambda_i = \sum_j M_{ji} I_j / N_i$ is the force of infection on age group $i$, $M_{ji}$ is the average number of people of age $i$ infected per unit time by an infectious individual of age $j$ in a fully susceptible population, and $\gamma$ and $\mu$ are constants representing to the inverse of the mean latent period and mean infectious period respectively. {\highlight The size of the recovered compartment $R_i$ may be obtained via the conservation equation $S_i+E_i+I_i+R_i=N_i$ and its dynamics do not need to be explicitly modelled.}

We defined the matrix $M$ to be the weighted sum of the household contact matrix $C^{(h)}$ and non-household contact matrix $C^{(n)}$:
\begin{equation}
    M = a_h C^{(h)} + a_n C^{(n)},
\end{equation}
where $a_h$ and $a_n$ are constants representing the infection rate per unit time of household or non-household contacts respectively. For the non-household matrix $C^{(n)}$ in all models, we used the sum of the school, work and other contact matrices estimated by \cite{prem2017projecting} for New Zealand, {\highlight after balancing via Eq. \eqref{eq:detbal}}. We investigated model behaviour over a range of values of $a_h$ and $a_n$. The basic reproduction number is given by
\begin{equation} \label{eq:R0}
R_0 = \rho(M^T)/\mu,    
\end{equation}
where $\rho(M^T)$ denotes the dominant eigenvalue of the transpose of $M$. We scaled the {\highlight balanced} Prem home contact matrix so that it had the same dominant eigenvalue as the contact matrix constructed from household composition data, and we also multiplied the {\highlight balanced} Prem non-household matrix by the same scaling factor. This ensured that we were comparing models with the same value of $R_0$.

\subsection{Agent-based epidemic model}  \label{sec:ABM}
The compartment-based model assumed that the population within each age group is well mixed. This ignored the effect of local contact network saturation on transmission dynamics. Fully accounting for the effects of network saturation is complex as it requires assumptions about the architecture of the overall social contact network and the relative frequency of transmission along different network edges \cite{kiss2017mathematics,harvey2020network}. However, we can investigate the effects of household saturation by considering an agent-based model that assigns individuals to specific households. 

To do this, we constructed a synthetic population from the household composition data. Each household type $k$ consists of $H_{ki}$ individuals of age group $i$ ($i=1,\ldots,8$) -- see Table \ref{tab:Stats NZ data}a). We represented a household of type $k$ in the synthetic population by creating an explicit list of individuals and their age groups in that household. We then repeated this list $f_k$ times. This resulted in a synthetic population of $N$ individuals, each with attributes representing their age group and household ID number  (see Table \ref{tab:Stats NZ data}b). 

We simulated epidemic dynamics in the synthetic population using a discrete-time agent-based model. At each daily time step $t$, each individual was in one of four states: susceptible, exposed, infectious or recovered. We assumed that each infectious individual had a daily probability $1-e^{-a_h \delta_t}$ of infecting each susceptible individual in the same household, where $\delta_t=1$ day is the time step. {\highlight This is a density-dependent model for within-household transmission, i.e. infection risk scales with the number of household members who are infectious rather than the proportion who are infectious. This is a reasonable assumption for infections that are transmitted via the respiratory route or other close contact \cite{house2009household}, but is not applicable in other situations such as sexually transmitted infections.  }

We modelled non-household infections by assuming that an infectious individual in age group $i$ at time $t$ {\highlight would have an infectious contact with a Poisson distributed number $N_\mathrm{inf}$ of individuals in age group $j$, with $N_\mathrm{inf}\sim\mathrm{Poiss}(a_n C^{(n)}_{ij} \delta_t$). The $N_\mathrm{inf}$ contacts were chosen at random} from the population of individuals in age group $j$ and those who were in the susceptible state at time $t$ were moved to the exposed state at time $t+\delta t$.

Individuals in the exposed state at time $t$ moved to the infectious state at time $t+\delta t$ with probability $\gamma \delta t$. Individuals in the infectious state at time $t$ moved to the recovered state at time $t+\delta t$ with probability $\mu\delta t$. We ran the model until there were no individuals remaining in either the exposed or the infectious state. 

We initialised both the compartment-based and agent-based models by assuming that a fraction $e_0$ of each age group (randomly selected in the case of the agent-based model) was in the exposed state at $t=0$, with all other individuals in the susceptible state. Parameter values used in the models are shown in Table \ref{tab:params}. {\highlight Note that because of susceptible depletion within households, the basic reproduction numbers for the compartment-based and agent-based model will not be the same.  }
 
The raw data, the code used to produce the results in this article, and an electronic version of the New Zealand household contact matrix and synthetic population are publicly available at: \url{https://github.com/michaelplanknz/household-contact-matrices-nz}. All analysis was carried out in {\em Matlab R2022b}.

\begin{table}
    \centering
    \begin{tabular}{ll}
    \hline
      Daily infection rate for household contacts   &  $a_h=0.025$--$0.065$ \\
      Daily infection rate for non-household contacts   &  $a_n=0.025$--$0.065$ \\
     Average latent period & $1/\gamma = 1$ day \\
     Average infectious period & $1/\mu = 4$ days \\
     Initial exposed fraction & $e_0=10^{-5}$ \\
         \hline
    \end{tabular}
    \caption{Parameter values used in the model. The daily infection rate parameters were chosen to give a range of values of the basic reproduction number $R_0$ {\highlight for the compartment-based ODE model (which was calculated according to Eq. \eqref{eq:R0})} between $1$ and $3$. The latent period and infectious period parameters are approximately representative of respiratory viruses such as SARS-CoV-2 or influenza {\highlight \cite{cowling2009estimation,park2023inferring}. However, it should be noted that these parameters only affect the timescale for the epidemic and do not change the age-specific attack rates.} The initial condition $e_0$ for the exposed fraction of the population does not have a substantial effect on model outputs.  }
    \label{tab:params}
\end{table}

\begin{figure}
    \centering
    \includegraphics[width=\textwidth]{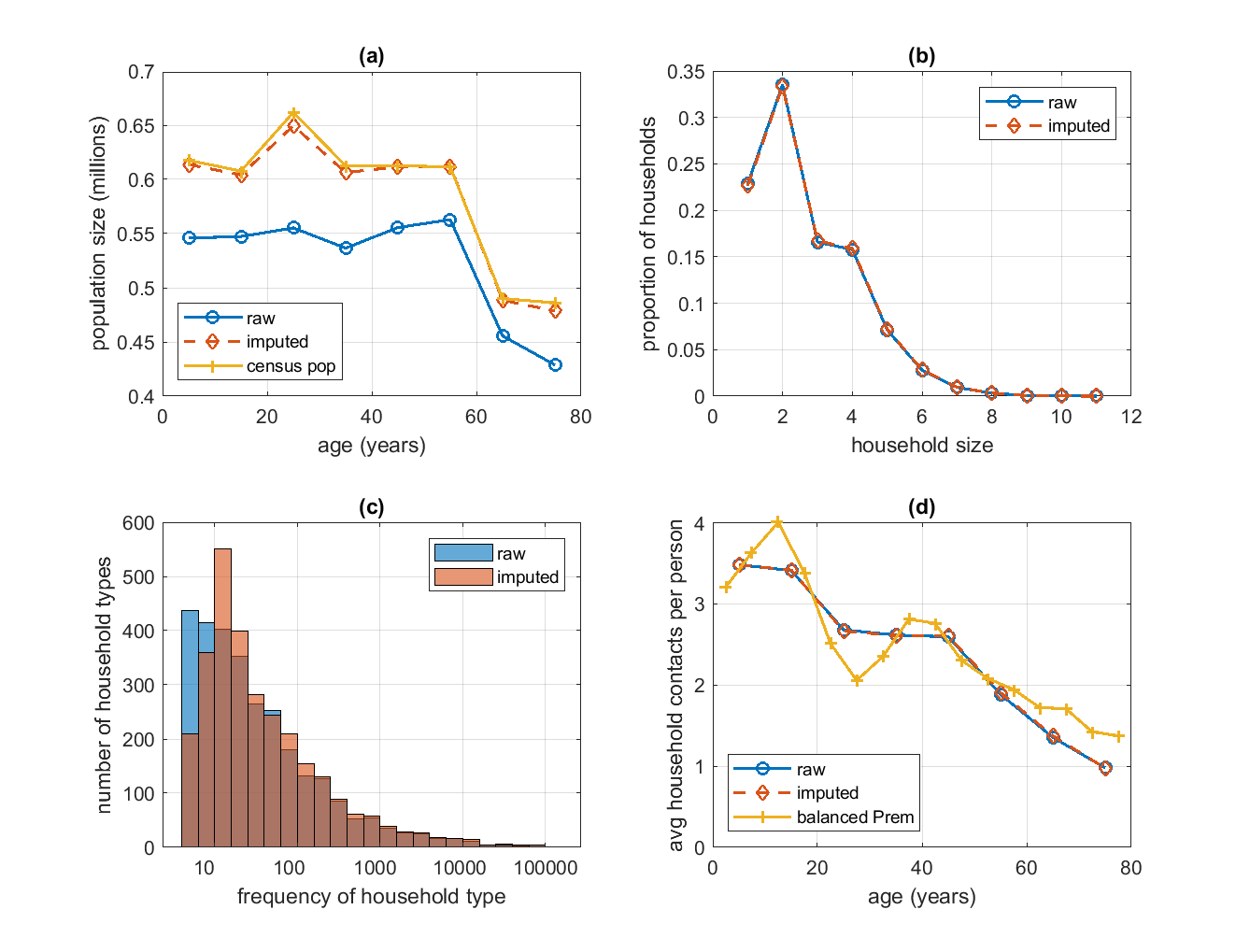}
    \caption{(a) Population size in 10-year age groups according to the household composition data before (blue) and after (red) imputation, and the 2018 Stats NZ census population (yellow). (b) Distribution of household size according to the household composition data before (blue) and after (red) imputation. (c) Number of household types that have a particular frequency in the household composition data before (blue) and after (red) imputation (e.g. the first bar says that before imputation there were around 420 distinct household types that had a frequency of less than 10 in the data). (d) The average total number of household contacts (of any age) per person according to the household composition data before imputation (blue) and after imputation (red), and according to the balanced Prem home contact matrix for New Zealand (yellow). In (a) and (d), all points are plotted at the midpoint of their age group. }
    \label{fig:age_graphs}
\end{figure}

\section{Results}

\subsection{Household size distribution and data imputation} 
A total of 4,300,107 individuals in 1,593,804 households were represented in the raw household data (mean household size $2.70$). After rounding and suppression of low counts, there were 4,187,691 people in 1,580,445 households of 2,908 distinct types (mean household size $2.65$), {\highlight meaning that 112,416 individuals ($2.6$\%) and 13,359 households ($0.8$\%) were suppressed (see Supplementary Table S1)}. The represented population corresponded to 89\% of the official 2018 census total population size estimate of 4,699,764 \cite{stats_nz_2019}, ranging from 84\% in the 20-30-year age group to 93\% in the 60-70-year age group. 

The imputation procedure added 476,553 individuals in 172,459 households. This resulted in a population size that was within $\pm 1.9\%$ of the census population in each 10-year age group (see Figure \ref{fig:age_graphs}a). The most common household size was 2 (approximately 33\% of all households), followed by 1, 3 and 4, with a sharp drop off in the frequency of households of size 5 and above (see Figure \ref{fig:age_graphs}b). The imputation procedure resulted in minimal change to household size distribution, with the most noticeable change being a slight decrease in the proportion of households of size 1. There was a modest change to the distribution of household type frequencies, with a reduction in the number of rare (count $<10$) household types and an increase in the number of mid-frequency (count 10--400) household types (Figure \ref{fig:age_graphs}c). {\highlight These changes resulted from rare households being disproportionately added to the population, meaning that they became less rare. However these changes affected a relatively small proportion of the overall household distribution.} Thus, the imputation procedure provided a reasonable balance between approximating the correct population age structure and preserving the distribution of household types.  

\begin{figure}
    \centering
    \includegraphics[width=\textwidth]{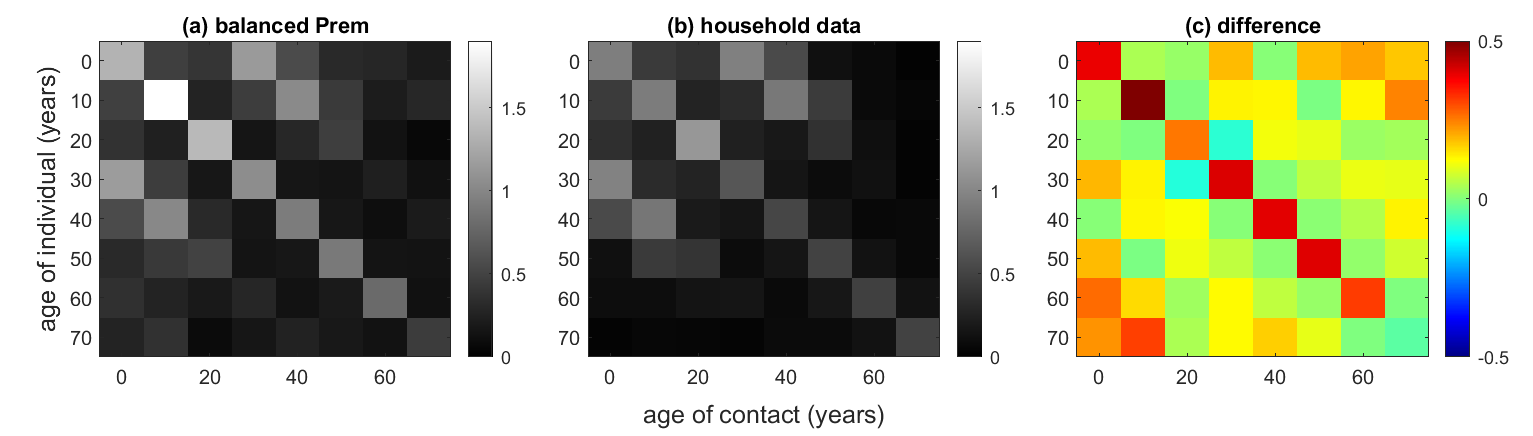}
    \caption{(a) The balanced Prem home matrix for New Zealand aggregated into 10-year age bands. (b) The contact matrix constructed from New Zealand household composition data after imputation. (c) The difference between the two matrices (a minus b). The $i^\mathrm{th}$ row and $j^\mathrm{th}$ column corresponds to the average number of household contacts that a person in age group $i$ has with individuals in age group $j$. }
    \label{fig:matrices}
\end{figure}

\subsection{Household contact matrices}
{\highlight The contact matrix constructed from the imputed household composition data shared some similarities with the balanced Prem home contact matrix} (Figure \ref{fig:matrices}). Both matrices featured strong diagonal bands, corresponding to household contacts in the same age group. They also had slightly weaker off-diagonal bands, corresponding to household contacts approximately 30 years apart, likely representing parents and children. 

However, there were some notable differences. {\highlight In the balanced Prem matrix, the total number of contacts per person exhibited peaks in the 10--15 and 35--40 year-old age groups before declining with age (Figure \ref{fig:age_graphs}d). In the matrix constructed from the household composition data, the total number of contacts per person declined monotonically with age and, above age 50 years, declined more steeply than in the balanced Prem matrix (Figure \ref{fig:age_graphs}d). }

{\highlight The balanced Prem matrix had a stronger diagonal, indicating more strongly age-assortative mixing, particularly in the 10--20-year age band (Figure \ref{fig:matrices}). It had slightly stronger off-diagonals, corresponding to first generation mixing. There was also some evidence of a weak secondary diagonal in the balanced Prem matrix, corresponding to second generation mixing, which appeared to be absent in the household composition data. Note the contact matrix constructed from the raw household composition data was almost identical to that constructed from the imputed data (see Supplementary Figure S1). }

Electronic versions of the household contact matrix and synthetic population are available in Supplementary Tables S4--S8. 

\begin{figure}
    \centering
    \includegraphics[width=\textwidth]{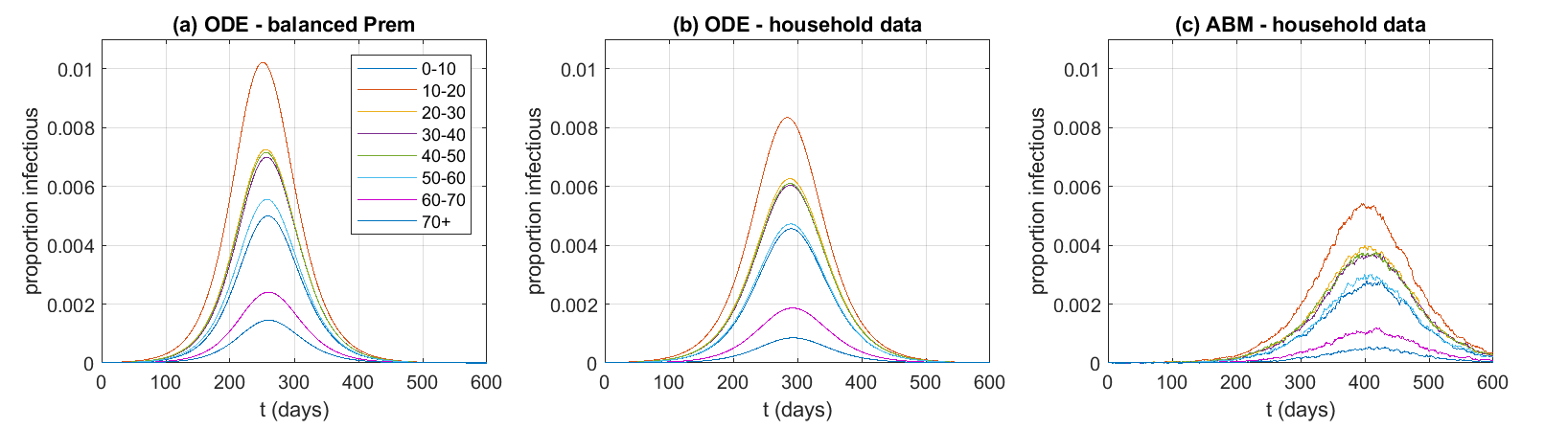}
    \caption{Epidemic curves showing the proportion $I_i(t)/N_i$ of each age group that is in the infectious state at time $t$ under the compartment-based ODE model with (a) the {\highlight balanced} Prem home contact matrix for New Zealand; (b) the household composition contact matrix; and (c) the agent-based model using the household composition data. Daily infection rate parameters for household contacts $a_h=0.025$ and non-household contacts $a_n=0.025$, which correspond to a basic reproduction number of $R_0=1.14$ {\highlight in the compartment-based ODE model. Note the model with the balanced Prem matrix (a) was run in 5-year age bands and then aggregated up to 10-year age bands for comparison with the models in (b,c) .} }
    \label{fig:epi_curves}
\end{figure}

\subsection{Epidemic dynamics}

When the basic reproduction number {\highlight in the compartment-based ODE model} is only moderately above the threshold value of $1$ ($R_0=1.14)$, the models using the household composition contact matrix and the {\highlight balanced} Prem matrix produced broadly similar results (Figure \ref{fig:epi_curves}a-b). The 10-20-year age group had the highest prevalence and older age groups experienced lower infection rates and peaked slightly later than younger groups. {\highlight There was a wider variability in peak prevalence with the balanced Prem matrix than with the household composition matrix. }
The agent-based model, which explicitly simulated transmission within individual households, produced an epidemic that was noticeably smaller and peaked later than the compartment-based models (Figure \ref{fig:epi_curves}c). This was expected because susceptible individuals within an infected household will tend to become depleted and household transmission chains will eventually self-extinguish. {\highlight Therefore, although the transmission rate parameters $a_h$ and $a_n$ are the same for both models, the basic reproduction number will generally be smaller for the agent-based model. }

\begin{figure}
    \centering
    \includegraphics[width=\textwidth, trim={0 0 1.95cm 0},clip]{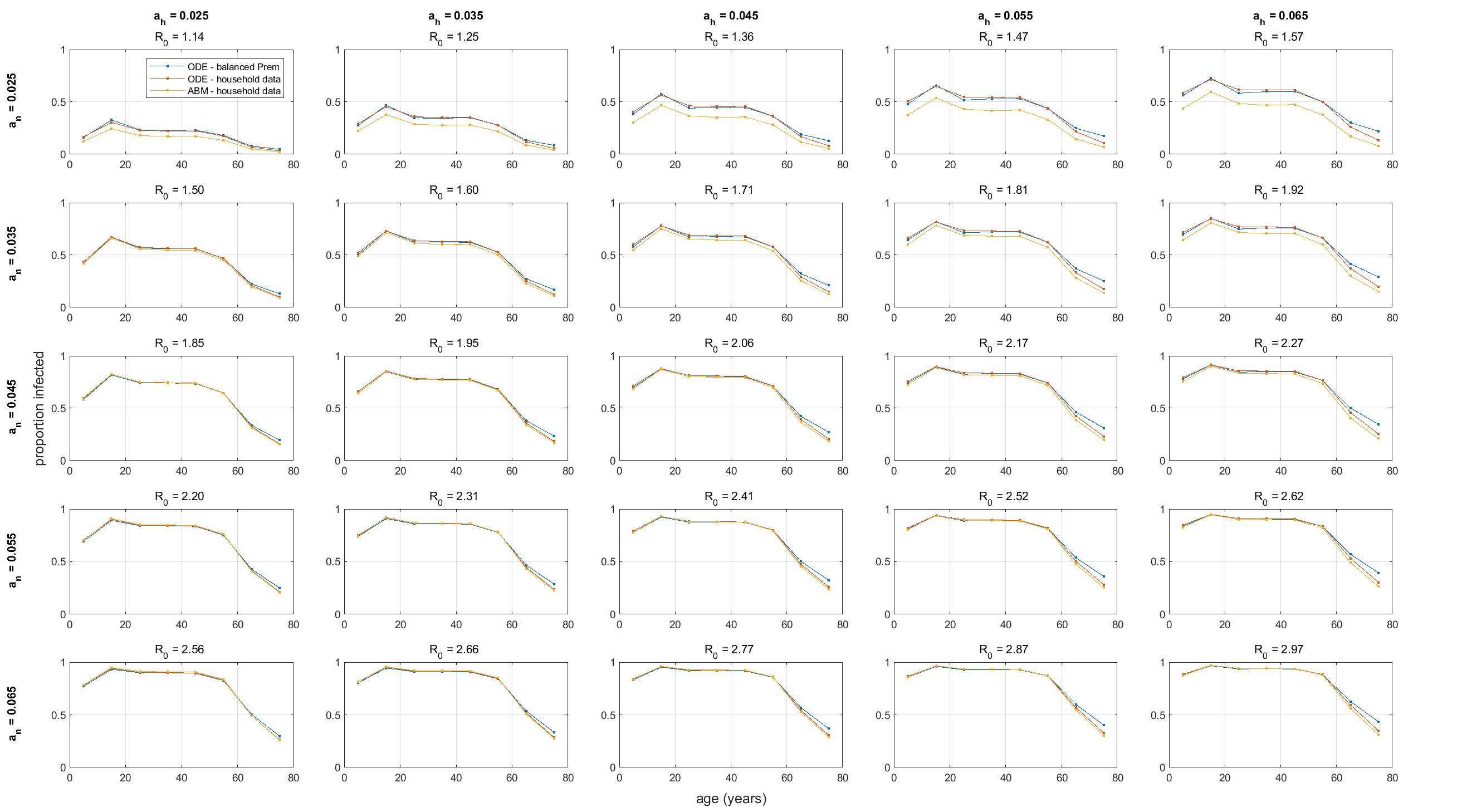}
    \caption{Attack rates for each age group for a range of values of the daily infection rate for household contacts ($a_h$) and non-household contacts ($a_n$) for: the compartment-based ODE model with the {\highlight balanced} Prem home contact matrix for New Zealand (blue); the household composition contact matrix (red); and the agent-based model using the household composition data (yellow). Each panel shows the value of the basic reproduction number $R_0$ {\highlight in the compartment-based ODE model} for that combination of values of $a_n$ and $a_h$. {\highlight Note the model with the balanced Prem matrix (blue) was run in 5-year age bands and then aggregated up to 10-year age bands for comparison with the other models.} All points are plotted at the midpoint of their age group. }
    \label{fig:attack_rates}
\end{figure}

The results in Figure \ref{fig:epi_curves} confirm that the qualitative behaviour of the models was as expected. We now turn to a systematic comparison of the age-dependent attack rate (i.e. proportion of each age group that became infected during the epidemic) for different values of the household and non-household infection rate parameters $a_h$ and $a_n$. {\highlight Prem et al. \cite{prem2017projecting} implicitly assumed $a_h$ and $a_n$ were equal (in their no-control scenario) by simply adding the location-specific contact matrices together. However, the relative values of $a_h$ and $a_n$ may be pathogen- and context-dependent. For example, if within-household contacts are more likely to involve prolonged periods of close contact and this increases the probability of transmission, then it would be reasonable to set $a_h>a_n$.  Figure \ref{fig:attack_rates} shows results for a range of both parameters which, while not necessarily covering all potential scenarios, is sufficient to show the broad trends in how age-specific attack rates depend on the relative strength of within versus between household transmission.} Across a range of values of these $a_h$ and $a_n$, the attack rate was highest around age 10--20 years and declined with age above around 40 years. This is consistent with the relationship between age and average number of contacts per person (Figure \ref{fig:age_graphs}d). {\highlight It is also qualitatively consistent with lower inferred infection rates in older age groups during the Omicron waves of Covid-19 in New Zealand \cite{lustig2023modelling}, although this cannot be validated directly because there is a lack of representative data on infections rates and self-reported case data are likely to be unrepresentative of the true age patterns.}

Increasing either $a_h$ or $a_n$ increased $R_0$ and therefore increased the attack rate and overall epidemic size. However, these two parameters had different effects in the different models considered. When $a_n$ was high relative to $a_h$, the three models behaved quite similarly (lower-left panels in Figure \ref{fig:attack_rates}). This is {\highlight as expected} because non-household transmission dominates when $a_n>a_h$ and all three models had the same assumptions about this transmission mode (using the {\highlight balanced} non-home Prem matrices). 

When $a_h$ was high relative to $a_n$, there were significant differences among the models (upper-right panels in Figure \ref{fig:attack_rates}). The agent-based model had consistently lower attack rates across all age groups due to depletion of susceptibles within households. The household composition contact matrix led to similar attack rates to the {\highlight balanced}  Prem matrix in under-60-year-olds, but consistently lower attack rates than the {\highlight balanced} Prem matrix in over-60-year-olds. This could be an important finding with implications for choice of control strategy, especially for pathogens with a strong age gradient in clinical severity. 

For visual simplicity, Figure \ref{fig:attack_rates} shows results from a single realisation of the imputation procedure and the agent-based model. To check the stochastic variability in attack rates, we calculated the median and 95\% range of the attack rates across $m=100$ independent realisations. We found that the amount of variation between realisations was relatively small (see Supplementary Figure S2). This is due to the fact that, although stochastic effects may be important in the early stages of an epidemic when the number of infections is small, the final epidemic size in a large population is relatively insensitive to this \cite{miller2012note}. It also confirmed that stochastic variability in the imputation procedure did not substantially impact results.  {\highlight Because our household contact matrix is constructed from census as opposed to survey data, it eliminates some of the uncertainty associated with survey sampling. However, we acknowledge that substantial uncertainty remains in the non-household matrix and that this leads to uncertainty in epidemic trajectories that is not captured here \cite{van2022learning}. }

The agent-based model additionally enables results to be stratified by household size. When household transmission dominated ($a_h>a_n$), there was an approximately linear relationship between household size and attack rate (see Supplementary Figure S3, upper-right panels). This was because there are more opportunities for the infection to enter larger households, and a lower probability that household transmission chains will stochastically self-extinguish before infecting a high proportion of household members. When non-household transmission dominated ($a_n>a_h$), the relationship between attack rate and household size was relatively flat for households of size $3$ or larger. However, there was a substantially lower attack rate for individuals living in households of size $1$ or $2$ (see Supplementary Figure S3, lower-left panels).

\section{Discussion}
In this article, we have presented data on the age group composition of households in Aotearoa New Zealand and developed methods for using these data to parameterise age-structured compartment-based and agent-based models of human infectious disease dynamics. For compartment-based models, we used the data to derive a contact matrix, which defines the average number of contacts a person in each group has with individuals in each other age group. For agent-based models, we constructed a synthetic population in which each individual is explicitly associated with a unique household. 

We compared these outputs to a home contact matrix that was projected onto the New Zealand population by \cite{prem2017projecting} from social survey data collected from European countries in 2005-2006 \cite{mossong2008social}. We found that individuals aged 10-19 years and individuals aged over 60 years had fewer household contacts than estimated in the projected synthetic contact matrix, whereas individuals aged 20-29 years had more household contacts than in the synthetic matrix. 

Although the qualitative behaviour of epidemic models using the New Zealand specific household data and the projected matrix of  \cite{prem2017projecting} were similar, we found that models based on the New Zealand data led to consistently lower attack rates in over-60-year-olds than using the Prem matrix. This could be an important difference from a public health perspective because, if there is a strong age gradient in clinical severity, the health impact will be highly sensitive to attack rates in older age groups. Having models that can estimate age-specific attack rates as accurately as possible is therefore crucial for impact assessment, e.g. for emerging pathogens and pandemic threats.  

We have provided a methodology that can be applied to updated data (e.g. from future census) or in other jurisdictions where comparable household composition data as available. By making the household contact matrices, synthetic population and associated metadata and code available in electronic form, our results may be used by other researchers to parameterise infectious disease models for the New Zealand population. By construction, our household contact matrix satisfies balance conditions \cite{arregui2018projecting} for the population to which it applies. We recommend that whenever contact matrices are estimated, the corresponding population age distribution is published alongside them. This would avoid the need to artificially impose balance conditions, which is necessary for contact matrices that were projected for a population with different, unknown age structure \cite{hamilton2024examining,steyn2022covid}.

We have explored models across a range of household and non-household transmission rate parameters. In general, these parameters will depend on the mode(s) of transmission and will be pathogen-specific \cite{longini1982household,bi2021insights}. They will also respond differentially to control measures and behavioural change. For example, social distancing behaviours, school or workplace closures, and case isolation measures would primarily reduce non-household transmission rate, and may not affect or may even increase the household transmission rate \cite{kwok2011modelling,paul2021characteristics}. We used fixed values of the mean latent period and infectious period, broadly representative of respiratory viruses such as influenza and SARS-CoV-2 \cite{cauchemez2004bayesian,cowling2009estimation,park2023inferring}. However, this is not restrictive because changing the latent or infectious period only changes the timescale on which the epidemic occurs without changing the overall age-specific attack rates.  

Our methods and results have several important limitations. The household composition data is incomplete, with approximately 11\% of the 2018 census usually resident population missing from the data. This may be because these individuals were not assigned to a household, or were assigned to a household that was suppressed from the released dataset due to low counts and preservation of confidentiality. This data is not missing at random, but is disproportionately weighted towards 20--29-year-olds, individuals living in more complex households (i.e., larger households and households containing extended families and more than one family), and households with M\=aori or Pacific residents \cite{stats_nz_adding-admin-records,stats_nz_2018_census_families_and_households}. Although we used an imputation method to recover the approximate size and age structure of the 2018 usually resident population, the imputed data will contain biases, which could affect model outputs. 

Our results describe an epidemic in a population with the same age structure as in the 2018 census. Ideally, models would use up-to-date household composition data, but this will not always be available as the census is only carried out every five years. To apply the models to a contemporaneous population, the imputation procedure could be modified to target the most recent estimated residential population.   

We have treated household structure as being static for the duration of the epidemic. In reality, household structure changes over time due to births, deaths and movement between households. This could impact infectious disease dynamics, particularly when the relevant time frames are long, for example epidemics of pathogens with a relatively long generation interval and diseases in an endemic or seasonal pattern \cite{glass2011incorporating}.

We have only considered New Zealand data on household contacts, and all our models used contact matrices derived from European data \cite{prem2017projecting} to parameterise non-household contacts. Whilst our agent-based model enables the effects of saturation of household transmission to be explored, it ignores the consequences of structure in the non-household contact network, e.g. local network saturation due to clustering of non-household contacts \cite{diekmann1998deterministic,keeling1999effects}.  
An important objective for future research is to use New Zealand-specific data to estimate and test contact matrices for contacts that occur outside the home. These could be derived from a diary-based survey study specifically designed for this purpose. Alternatively, they could be approximated from other sources of information such as education and employment data in the Stats NZ Integrated Data Infrastructure \cite{statsnz_idi}, see e.g. \cite{turnbull2022investigating,harvey2023summary}.

We have only considered models of an epidemic in a closed population, where infection confers lasting immunity. The models could be extended to cover a wider range of situations, such as births and deaths, and waning immunity. These processes generally lead to models with an endemic equilibrium \cite{anderson1991infectious}, and a natural question is how sensitive is the age-specific equilibrium prevalence to different methods for parameterising age-specific contact patterns. We leave this as a question for future work.

\subsection*{Acknowledgements}
This work includes customised Stats NZ data which are licensed by Stats NZ for re-use under the Creative Commons Attribution 4.0 International licence.
All data and other material produced by Stats NZ constitutes Crown copyright administered by Stats NZ. The authors are grateful to two anonymous reviewers for helpful comments on a previous version of this manuscript.


\begin{thebibliography}{10}

\bibitem{verity2020estimates}
Verity R, Okell LC, Dorigatti I, Winskill P, Whittaker C, Imai N, et~al.
\newblock Estimates of the severity of coronavirus disease 2019: a model-based analysis.
\newblock Lancet Infectious Diseases. 2020;20(6):669-77.

\bibitem{mcdonald2023inference}
McDonald SA, Teirlinck AC, Hooiveld M, van Asten L, Meijer A, de~Lange M, et~al.
\newblock Inference of age-dependent case-fatality ratios for seasonal influenza virus subtypes {A} ({H3N2}) and {A} ({H1N1}) pdm09 and {B} lineages using data from the {N}etherlands.
\newblock Influenza and Other Respiratory Viruses. 2023;17(6):e13146.

\bibitem{castle2000clinical}
Castle SC.
\newblock Clinical relevance of age-related immune dysfunction.
\newblock Clinical Infectious Diseases. 2000;31(2):578-85.

\bibitem{muller2021age}
M{\"u}ller L, Andr{\'e}e M, Moskorz W, Drexler I, Walotka L, Grothmann R, et~al.
\newblock Age-dependent immune response to the {BioNTech}/{Pfizer} {BNT162b2} coronavirus disease 2019 vaccination.
\newblock Clinical Infectious Diseases. 2021;73(11):2065-72.

\bibitem{anderson1985vaccination}
Anderson RM, May RM.
\newblock Vaccination and herd immunity to infectious diseases.
\newblock Nature. 1985;318(6044):323-9.

\bibitem{delvalle2013mathematical}
Del~Valle SY, Hyman JM, Chitnis N.
\newblock Mathematical models of contact patterns between age groups for predicting the spread of infectious diseases.
\newblock Mathematical Biosciences and Engineering. 2013;10:1475-97.

\bibitem{riley2007large}
Riley S.
\newblock Large-scale spatial-transmission models of infectious disease.
\newblock Science. 2007;316(5829):1298-301.

\bibitem{cauchemez2004bayesian}
Cauchemez S, Carrat F, Viboud C, Valleron AJ, Bo{\"e}lle P.
\newblock A Bayesian {MCMC} approach to study transmission of influenza: application to household longitudinal data.
\newblock Statistics in Medicine. 2004;23(22):3469-87.

\bibitem{kwok2011modelling}
Kwok KO, Leung GM, Riley S.
\newblock Modelling the proportion of influenza infections within households during pandemic and non-pandemic years.
\newblock PLoS One. 2011;6(7):e22089.

\bibitem{paul2021characteristics}
Paul LA, Daneman N, Brown KA, Johnson J, van Ingen T, Joh E, et~al.
\newblock Characteristics associated with household transmission of Severe Acute Respiratory Syndrome Coronavirus 2 ({SARS-CoV-2}) in {Ontario}, {Canada}: a cohort study.
\newblock Clinical Infectious Diseases. 2021;73(10):1840-8.

\bibitem{house2009household}
House T, Keeling MJ.
\newblock Household structure and infectious disease transmission.
\newblock Epidemiology and Infection. 2009;137(5):654-61.

\bibitem{hilton2019incorporating}
Hilton J, Keeling MJ.
\newblock Incorporating household structure and demography into models of endemic disease.
\newblock Journal of The Royal Society Interface. 2019;16(157):20190317.

\bibitem{keeling2011modeling}
Keeling MJ, Rohani P.
\newblock Modeling infectious diseases in humans and animals.
\newblock Princeton University Press; 2011.

\bibitem{schenzle1984age}
Schenzle D.
\newblock An age-structured model of pre-and post-vaccination measles transmission.
\newblock Mathematical Medicine and Biology: A Journal of the IMA. 1984;1(2):169-91.

\bibitem{diekmann2010construction}
Diekmann O, Heesterbeek JAP, Roberts MG.
\newblock The construction of next-generation matrices for compartmental epidemic models.
\newblock Journal of the royal society interface. 2010;7(47):873-85.

\bibitem{mossong2008social}
Mossong J, Hens N, Jit M, Beutels P, Auranen K, Mikolajczyk R, et~al.
\newblock Social contacts and mixing patterns relevant to the spread of infectious diseases.
\newblock PLoS Medicine. 2008;5(3):e74.

\bibitem{prem2017projecting}
Prem K, Cook AR, Jit M.
\newblock Projecting social contact matrices in 152 countries using contact surveys and demographic data.
\newblock PLoS Computational Biology. 2017;13(9):e1005697.

\bibitem{prem2021projecting}
Prem K, {van Zandvoort} K, Klepac P, Eggo RM, Davies NG, {Centre for the Mathematical Modelling of Infectious Diseases COVID-19 Working Group}, et~al.
\newblock Projecting contact matrices in 177 geographical regions: an update and comparison with empirical data for the {COVID-19} era.
\newblock PLoS Computational Biology. 2021;17(7):e1009098.

\bibitem{conmat}
Tierney N, Golding N, Babu A, Lydeamore M. Conmat: Builds contact matrices using {GAM}s and population data; 2024.
\newblock R package version 0.0.2.9000.
\newblock Available from: \url{https://idem-lab.github.io/conmat/dev/index.html}.

\bibitem{moore2021modelling}
Moore S, Hill EM, Dyson L, Tildesley MJ, Keeling MJ.
\newblock Modelling optimal vaccination strategy for {SARS-CoV-2} in the {UK}.
\newblock PLoS Computational Biology. 2021;17(5):e1008849.

\bibitem{bubar2021model}
Bubar KM, Reinholt K, Kissler SM, Lipsitch M, Cobey S, Grad YH, et~al.
\newblock Model-informed {COVID}-19 vaccine prioritization strategies by age and serostatus.
\newblock Science. 2021;371(6532):916-21.

\bibitem{ram_schaposnik_2021}
Ram V, Schaposnik LP.
\newblock A modified age-structured SIR model for COVID-19 type viruses.
\newblock Scientific Reports. 2021;11(1).

\bibitem{nguyen2021covid}
Nguyen T, Adnan M, Nguyen BP, de~Ligt J, Geoghegan JL, Dean R, et~al.
\newblock {COVID}-19 vaccine strategies for {Aotearoa New Zealand}: a mathematical modelling study.
\newblock The Lancet Regional Health--Western Pacific. 2021;15:100256.

\bibitem{kimathi2021age}
Kimathi M, Mwalili S, Ojiambo V, Gathungu DK.
\newblock Age-structured model for {COVID-19}: Effectiveness of social distancing and contact reduction in {Kenya}.
\newblock Infectious Disease Modelling. 2021;6:15-23.

\bibitem{conway2023covid}
Conway E, Walker CR, Baker C, Lydeamore MJ, Ryan GE, Campbell T, et~al.
\newblock COVID-19 vaccine coverage targets to inform reopening plans in a low incidence setting.
\newblock Proceedings of the Royal Society B. 2023;290(2005):20231437.

\bibitem{gimma2022changes}
Gimma A, Munday JD, Wong KL, Coletti P, van Zandvoort K, Prem K, et~al.
\newblock Changes in social contacts in England during the {COVID}-19 pandemic between {M}arch 2020 and {M}arch 2021 as measured by the {CoMix} survey: A repeated cross-sectional study.
\newblock PLoS Medicine. 2022;19(3):e1003907.

\bibitem{golding2023modelling}
Golding N, Price DJ, Ryan G, McVernon J, McCaw JM, Shearer FM.
\newblock A modelling approach to estimate the transmissibility of {SARS-CoV-2} during periods of high, low, and zero case incidence.
\newblock eLife. 2023;12:e78089.

\bibitem{vattiato2022assessment}
Vattiato G, Maclaren O, Lustig A, Binny RN, Hendy SC, Plank MJ.
\newblock An assessment of the potential impact of the {Omicron} variant of {SARS-CoV-2} in {Aotearoa New Zealand}.
\newblock Infectious Disease Modelling. 2022;7:94-105.

\bibitem{lustig2023modelling}
Lustig A, Vattiato G, Maclaren O, Watson LM, Datta S, Plank MJ.
\newblock Modelling the impact of the {O}micron {BA.5} subvariant in {N}ew {Z}ealand.
\newblock Journal of the Royal Society Interface. 2023;20(199):20220698.

\bibitem{datta2024impact}
Datta S, Vattiato G, Maclaren OJ, Hua N, Sporle A, Plank MJ.
\newblock The impact of {Covid}-19 vaccination in {Aotearoa New Zealand}: a modelling study.
\newblock Vaccine. 2024;42:1383-91.

\bibitem{steyn2022covid}
Steyn N, Plank MJ, Binny RN, Hendy SC, Lustig A, Ridings K.
\newblock A {COVID}-19 vaccination model for {Aotearoa New Zealand}.
\newblock Scientific Reports. 2022;12(1):1-11.

\bibitem{stats_nz_2018-census-population-and-dwelling-counts}
{Stats NZ}. 2018 Census population and dwelling counts; 2019.
\newblock Available from: \url{www.stats.govt.nz/information-releases/2018-census-population-and-dwelling-counts}.

\bibitem{stats_nz_adding-admin-records}
{Stats NZ}. Overview of statistical methods for adding admin records to the 2018 Census dataset; 2019.
\newblock Available from: \url{www.stats.govt.nz/methods/overview-of-statistical-methods-for-adding-admin-records-to-the-2018-census-dataset}.

\bibitem{stats_nz_2018_census_families_and_households}
{Stats NZ}. Families and households in the 2018 Census: Data sources, family coding, and data quality; 2019.
\newblock Available from: \url{www.stats.govt.nz/methods/families-and-households-in-the-2018-census-data-sources-family-coding-and-data-quality}.

\bibitem{stats_nz_2019}
{Stats NZ}. 2018 Census table finder -- Table 6; 2019.
\newblock Available from: \url{www.stats.govt.nz/tools/2018-census-table-finder/}.

\bibitem{arregui2018projecting}
Arregui S, Aleta A, Sanz J, Moreno Y.
\newblock Projecting social contact matrices to different demographic structures.
\newblock PLoS Computational Biology. 2018;14(12):e1006638.

\bibitem{hamilton2024examining}
Hamilton MA, Knight J, Mishra S.
\newblock Examining the influence of imbalanced social contact matrices in epidemic models.
\newblock American Journal of Epidemiology. 2024;193(2):339-47.

\bibitem{diekmann2000mathematical}
Diekmann O, Heesterbeek JAP.
\newblock Mathematical epidemiology of infectious diseases: model building, analysis and interpretation.
\newblock 5th ed. Wiley; 2000.

\bibitem{kiss2017mathematics}
Kiss IZ, Miller JC, Simon PL.
\newblock Mathematics of epidemics on networks: from exact to approximate moels.
\newblock Springer; 2017.

\bibitem{harvey2020network}
Harvey E, Maclaren O, O'Neale D, Ortiz-Cervantes A, Patten-Elliott F, Turnbull S, et~al.
\newblock Network-based simulations of re-emergence and spread of {COVID}-19 in {A}otearoa {N}ew {Z}ealand.
\newblock Covid-19 Modelling Aotearoa. 2020.
\newblock Available from: \url{www.covid19modelling.ac.nz/simulations-of-re-emergence-and-spread/}.

\bibitem{cowling2009estimation}
Cowling BJ, Fang VJ, Riley S, Peiris JSM, Leung GM.
\newblock Estimation of the serial interval of influenza.
\newblock Epidemiology. 2009;20(3):344-7.

\bibitem{park2023inferring}
Park SW, Sun K, Abbott S, Sender R, Bar-On YM, Weitz JS, et~al.
\newblock Inferring the differences in incubation-period and generation-interval distributions of the {Delta} and {Omicron} variants of {SARS-CoV-2}.
\newblock Proceedings of the National Academy of Sciences. 2023;120(22):e2221887120.

\bibitem{miller2012note}
Miller JC.
\newblock A note on the derivation of epidemic final sizes.
\newblock Bulletin of Mathematical Biology. 2012;74(9):2125-41.

\bibitem{van2022learning}
van~der Vegt SA, Dai L, Bouros I, Farm HJ, Creswell R, Dimdore-Miles O, et~al.
\newblock Learning transmission dynamics modelling of COVID-19 using comomodels.
\newblock Mathematical Biosciences. 2022;349:108824.

\bibitem{longini1982household}
Longini~Jr IM, Koopman JS.
\newblock Household and community transmission parameters from final distributions of infections in households.
\newblock Biometrics. 1982:115-26.

\bibitem{bi2021insights}
Bi Q, Lessler J, Eckerle I, Lauer SA, Kaiser L, Vuilleumier N, et~al.
\newblock Insights into household transmission of {SARS-CoV-2} from a population-based serological survey.
\newblock Nature Communications. 2021;12(1):3643.

\bibitem{glass2011incorporating}
Glass K, McCaw JM, McVernon J.
\newblock Incorporating population dynamics into household models of infectious disease transmission.
\newblock Epidemics. 2011;3(3-4):152-8.

\bibitem{diekmann1998deterministic}
Diekmann O, De~Jong MCM, Metz JAJ.
\newblock A deterministic epidemic model taking account of repeated contacts between the same individuals.
\newblock Journal of Applied Probability. 1998;35(2):448-62.

\bibitem{keeling1999effects}
Keeling MJ.
\newblock The effects of local spatial structure on epidemiological invasions.
\newblock Proceedings of the Royal Society of London Series B: Biological Sciences. 1999;266(1421):859-67.

\bibitem{statsnz_idi}
{Stats NZ}. Integrated Data Infrastructure; 2024.
\newblock Available from: \url{www.stats.govt.nz/integrated-data/integrated-data-infrastructure/}.

\bibitem{turnbull2022investigating}
Turnbull SM, Hobbs M, Gray L, Harvey E, Scarrold WML, O'Neale DRJ.
\newblock Investigating the transmission risk of infectious disease outbreaks through the {Aotearoa} Co-incidence Network ({ACN}): a population-based study.
\newblock Lancet Regional Health-Western Pacific. 2022;20:100351.

\bibitem{harvey2023summary}
Harvey E, Hobbs M, Kvalsvig A, Mackenzie F, O'Neale D, Turnbull S.
\newblock A summary of risk factors for {C}OVID-19 infection in {A}otearoa {N}ew {Z}ealand.
\newblock Covid-19 Modelling Aotearoa pre-print. 2023.
\newblock Available from: \url{www.covid19modelling.ac.nz/a-summary-of-risk-factors-for-covid-19-infection-in-aotearoa-new-zealand/}.

\bibitem{anderson1991infectious}
Anderson RM, May RM.
\newblock Infectious diseases of humans: dynamics and control.
\newblock Oxford University Press; 1991.

\end{thebibliography}
\end{document}